\documentclass[aps,prb,amsmath,amssymb,superscriptaddress,twocolumn,showpacs]{revtex4}
\usepackage{graphicx}
\newcommand{\bk}{{\mathbf k}}
\newcommand{\bp}{{\mathbf p}}
\newcommand{\bQ}{{\mathbf Q}}
\newcommand{\barQ}{\overline{\mathbf Q}}
\newcommand{\bq}{{\mathbf q}}
\newcommand{\br}{{\mathbf r}}

\newcommand{\psidag}{\psi^{\dagger}}

\newcommand{\hx}{{\hat{x}}}
\newcommand{\hy}{{\hat{y}}}

\begin{document}

\title{Symmetry of the charge density wave in cuprates}

\author{Ashot Melikyan}
\affiliation{Materials Science Division, Argonne National Laboratory, Argonne, IL, 60439}
\affiliation{American Physical Society, Ridge, NY 11961}

\author{M. R. Norman}
\email{norman@anl.gov}
\affiliation{Materials Science Division, Argonne National Laboratory, Argonne, IL, 60439}

\date{\today}

\begin{abstract}
We derive and analyze an effective Ginzburg-Landau (GL) functional 
for a charge density wave (CDW) for a model of electrons on a tight binding square lattice
with density-density interactions.
We show, using realistic electronic dispersions for the cuprates, that for the simplest
GL theory, the preferred symmetry
is typically uni-directional (stripe) type, but inclusion of third-order terms tends
to destabilize this in favor of a checkerboard pattern
depending on the strength and range of the interaction.
This is of interest given the recent observation of such charge order in underdoped YBa$_2$Cu$_3$O$_{6+x}$.
\end{abstract}
\maketitle

\section{Introduction}
\label{section:Introduction}

Modulations of the local density of states in cuprates, which were observed by
scanning tunneling microscopy in Bi$_2$Sr$_2$CaCu$_2$O$_{8+\delta}$ 
(Bi2212) \cite{hoffman1,hoffman2,howald,vershinin,kohsaka2,parker} 
and Ca$_{2-x}$Na$_x$CuO$_2$Cl$_2$ \cite{hanaguri,kohsaka1}, have attracted significant attention.
Although a number of theories for these observations have been
proposed,
\cite{hdchen2002, hdchen2004, tesanovic2004, melikyan2004, 
balents2005, peregbarnea, kivelsonrmp, robertson, bena, ghosal,
hdchendwave}
the nature of the modulated state is still debated.
The modulations are strongest in the underdoped region of the phase diagram,
a faithful description of which could be a difficult task.  Fluctuations of the
superconducting order parameter, intrinsic disorder,
and competing/coexisting order, all potentially play a role in underdoped samples.

Nevertheless, short of knowing the exact Hamiltonian governing the
low energy phenomenology of the cuprates, it is instructive to isolate the role
of separate contributions by focusing on one of them at a time.
The goal of this article is to study the influence of 
the electronic dispersion in determining the nature of 
potential charge density waves in cuprates.
This is realized by deriving a Ginzburg-Landau (GL) free energy and analyzing  the
symmetry of the possible charge modulations,
with the coefficients of the free energy determined from dispersions extracted
from angle-resolved photoemission data. 

\section{Ginzburg-Landau free energy for a CDW}
Our starting point is the generalized extended Hubbard model of interacting electrons on
a tight binding square lattice. In terms of electron creation operators $\psidag_{\bk,\sigma}$,
the  Hamiltonian of the model is:
\begin{equation}
\label{Hubbard}
H = \sum_{\bk,\sigma} (\xi_{\bk}-\mu)\psidag_{\bk\sigma} \psi_{\bk\sigma} +
\frac12\sum_{\br, \br'}
g(\br-\br') n(\br) n(\br')
\end{equation}
where $\xi_{\bk}$ denotes the energy dispersion
(with $\mu$ the chemical potential), $\br$ and $\br'$ are
the sites of the lattice, and $n(\br)$ is the charge density. 
The model reduces to the Hubbard model when 
$g(\br-\br') = g_0\delta_{\br,\br'}$, and
to the so-called $U$-$V$ (extended Hubbard) model when $g(\br-\br')$ is also
non zero for the nearest neighbor sites $\br' = \br \pm a \hx, \br\pm a \hy$.
\begin{figure}
\includegraphics[width=0.9\columnwidth,clip]{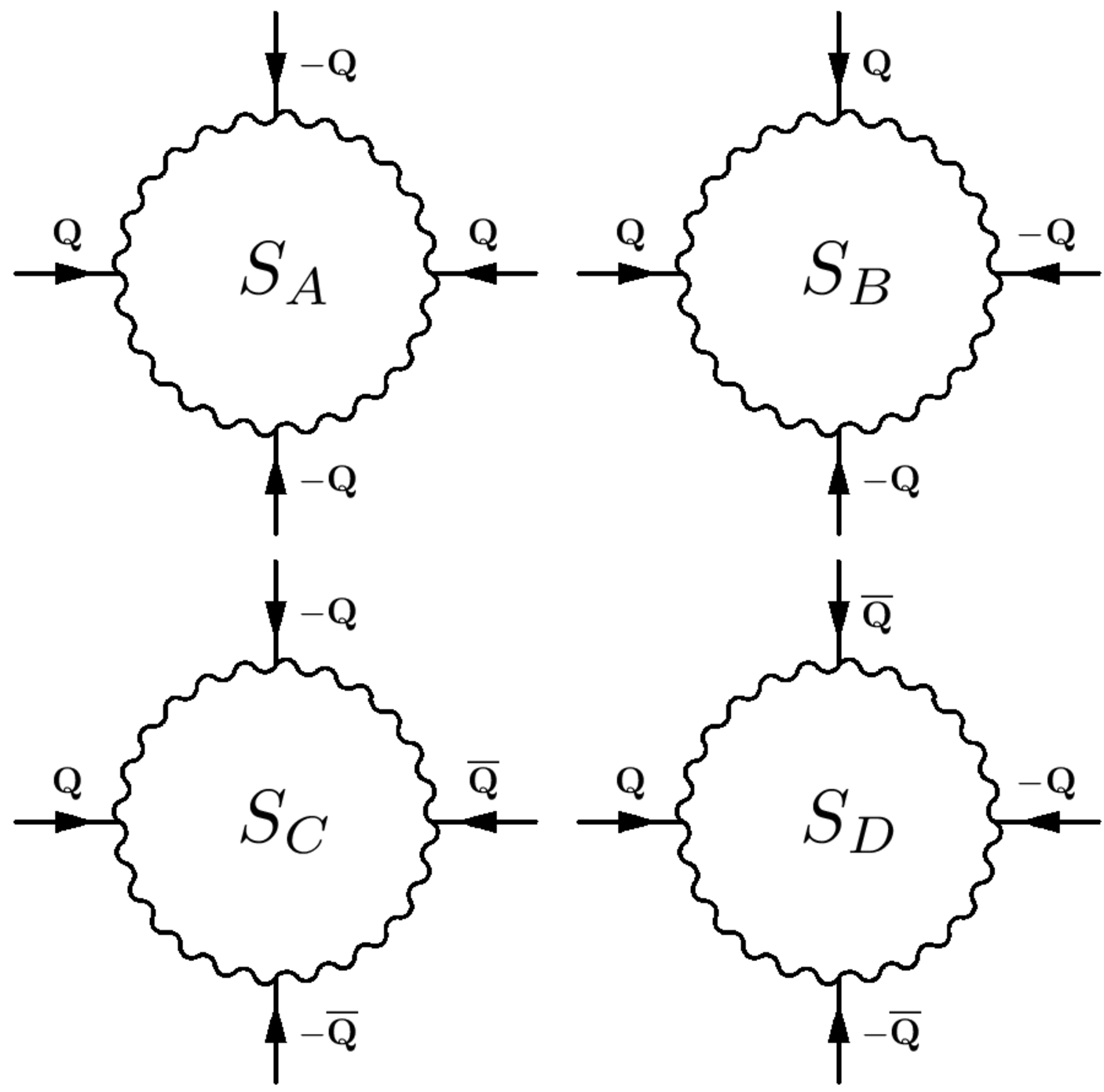}
\caption{Diagrams corresponding to the GL coefficients $S_A$,
$S_B$, $S_C$, and $S_D$.}
\label{Si}
\end{figure}
As a function  of the charge-density-wave order parameter $\Delta(\bQ)$,
the effective GL free energy (per lattice site) to quartic order can be written as
\begin{multline}
F({\Delta})-F(0)= \sum_{\bq,\Omega}|\Delta_{\bq,\Omega}|^2
\Biggl(
-\frac1{2g_{\bq}}-\chi(\bq,\Omega)\Biggr)
\\
-\frac23
\sum_{q_1,q_2,q_3}
\Delta_{q_1}
\Delta_{q_2}
\Delta_{q_3}\delta\left(\sum_{n=1}^3 q_n\right)
S_3(q_1,q_2,q_3)
\\
+\frac12
\sum_{q_1,q_2,q_3,q_4}
\Delta_{q_1}
\Delta_{q_2}
\Delta_{q_3}
\Delta_{q_4}
\delta\left(\sum_{n=1}^4 q_n\right)
S_4(q_1,q_2,q_3,q_4)
\end{multline}
where the summations are performed over the momenta $\bq_n$ and
the bosonic Matsubara frequencies $\Omega_n= 2\pi n/\beta$,
$F(0)$ is the free energy of the normal state ($\Delta_q=0$),
and the notation $q_n = (\bq_n, \Omega_n)$ is used for brevity. 
The coefficients of the expansion can be expressed through Greens
functions $G_{\bk, \omega_n} = (i\omega_n-\xi_{\bk}+\mu)^{-1}$ as
\begin{align}
\label{defS}
\chi(\bq,\Omega)&=-\frac1{N\beta}\sum_{k}
G_kG_{k+q}\\
S_3(\bq_i, \Omega_i)&=\frac1{N\beta}\sum_{k} G_{k} G_{k+q_1}
G_{k+q_1+q_2}\\
S_4(\bq_i,\Omega_i)&=\frac1{N\beta}\sum_{k}
G_{k} 
G_{k+q_1}
G_{k+q_1+q_2}
G_{k+q_1+q_2+q_3}
\end{align}
where $k=(\bk, \omega_n)$.
After  summation over fermionic frequencies $\omega_n = \pi(2n+1)/\beta$,
one obtains the usual result for the (2D) charge susceptibility
\begin{equation}
\chi(\bq,\Omega)
= -  \int \frac{a^2d\bk}{(2\pi)^2}
\frac{f(\xi_{\bk})-f(\xi_{\bk+\bq})}{\xi_{\bk}-\xi_{\bk+\bq}+i\Omega}
\end{equation}
with the integral over the first Brillouin zone.
In our mean field theory, we only need the cubic and quartic coefficient functions $S_3$ and $S_4$ 
for $\Omega_i=0$
\begin{equation}
S_3(\bp_i)=\int \frac{a^2d\bk}{(2\pi)^2}
\left[\frac{f(\xi_{\bk})}{(\xi_{\bk}-\xi_{\bk+\bp_1})(\xi_{\bk}-\xi_{\bk+\bp_1+\bp_2})}
+c.p.\right]
\end{equation}
where $c.p.$ denotes the two other terms obtained by cyclic permutation of
the momenta $(\bk,\bk+\bp_1,\bk+\bp_1+\bp_2)$. Similarly, for $S_4$
one finds
\begin{multline}
\label{S_4_summed_generalcase}
S_4(\bp_i)= \int \frac{a^2d\bk}{(2\pi)^2}
\Biggl(f(\xi_{\bk})(\xi_{\bk}-\xi_{\bk+\bp_1})^{-1}\\
(\xi_{\bk}-\xi_{\bk+\bp_1+\bp_2})^{-1}(\xi_{\bk}-\xi_{\bk+\bp_1+\bp_2+\bp_3})^{-1}+c.p.\Biggr)
\end{multline}
where $c.p.$ denotes the three other terms obtained by cyclic permutation of
the momenta $(\bk,\bk+\bp_1,\bk+\bp_1+\bp_2,\bk+\bp_1+\bp_2+\bp_3)$.
\begin{figure*}
\includegraphics[width=\textwidth,clip]{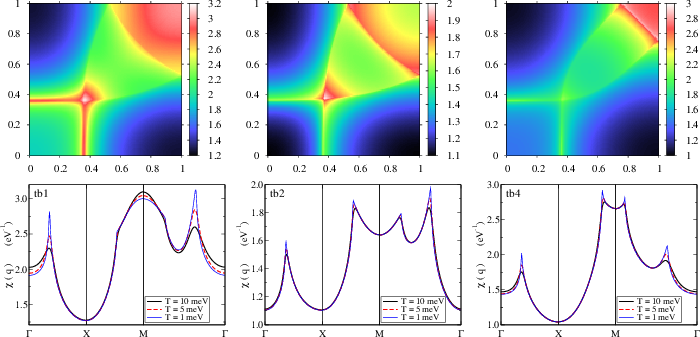}\\
\caption{\label{Fig:colormap_chi} (Color online) $\chi(\bq)$ calculated from the tight binding dispersions
tb1, tb2, and tb4 (left to right) \cite{norman2007}. Top row: Two-dimensional map of
$\chi(\bq)$ for $T=0.001t$, where $t$ is the near neighbor hopping parameter of the
tight binding fit for $\xi_{\bk}$.  Since $\chi(\bq)$ is four-fold symmetric,
only one quarter of the Brillouin zone is shown. Bottom row: $\chi(\bq)$ calculated
for $T=1$ meV, $5$ meV, and $10$ meV along the various symmetry lines of the zone, with
$\Gamma= (0,0)$, $X = (1,0)$, and $M = (1,1)$ in $\pi/a$ units. }
\end{figure*}
As the temperature or the effective interaction
$g(\br-\br')$ varies, for a specific set of equivalent momenta $\bQ_i$, the
quadratic coefficient of the GL expansion may change sign, and a phase
transition takes place. In the vicinity of such a transition,
it is sufficient to consider only the coefficients of the GL free energy evaluated at
momenta $\bQ_i$ and their harmonics.  We will be specifically  interested in
situations where these equivalent momenta lie along the symmetry lines
of the Brillouin zone, and thus the total number of them  is restricted to four for the square lattice
(typically, the maxima of $\chi$ will lie along such lines).
Consider first a simplified expansion where only the fundamental
harmonics at $\bQ_i$ are retained. In such a  case, the GL expansion reduces to
\begin{multline}
\label{Seff}
F(\Delta)-F(0) =
r_{\bQ}(|\Delta_{\bQ}|^2+|\Delta_{\barQ}|^2)
+\\
\frac12(2S_A+4S_B)
\Biggl(2\gamma|\Delta_{\bQ}|^2|\Delta_{\barQ}|^2+
|\Delta_{\bQ}|^4+|\Delta_{\barQ}|^4
\Biggr)
\end{multline}
where $\barQ\perp \bQ$,
\begin{equation}
\label{rQ}
r_{\bQ} = - g_{\bQ}^{-1}-2\chi(\bQ),
\end{equation}
$\Omega_i=0$ is implied,
and the coefficients $S_{A,B,C,D}$, which are described by the diagrams in Fig.~\ref{Si}, are defined
as
\begin{equation}
\label{SABCD}
\begin{aligned}
S_A&= S_4(\bQ,-\bQ,\bQ,-\bQ)\\
S_B&= S_4(\bQ,\bQ,-\bQ,-\bQ)\\
S_C&= S_4(\bQ,-\bQ,\barQ,-\barQ)\\
S_D&= S_4(\bQ,\barQ,-\bQ,-\barQ)
\end{aligned}
\end{equation}
with $\gamma=\frac{8S_C+4S_D}{2S_A+4S_B}$.
Formally identical GL expansions have been studied in the past,
and for a momentum-independent interaction  have been recently
employed by Yao {\it et al.}  in connection with CDWs that occur
in  the rare earth tri-tellurides (Ref.~\onlinecite{hongkivelson}).
The stability requirement dictates that $2S_A+4S_B>0$ and $\gamma>-1$. Provided these
are satisfied, the symmetry of the 
CDW in the ordered phase is determined \cite{robertson}  by $\gamma$: when $\gamma>1$, the free
energy is minimized by choosing either $\Delta_{\bQ} = 0$ or $\Delta_{\barQ} = 0$
resulting in one-dimensional stripes. For $\gamma < 1$, the minimum of the GL free
energy is achieved by choosing $|\Delta_{\bQ}|=|\Delta_{\barQ}|$, and the CDW
has a two-dimensional ``checkerboard'' pattern.

\section{Charge susceptibility and instability wave vectors in  cuprates}
The calculations were performed for three sets of six-parameter
tight-binding fits to the dispersion, based
on angle-resolved photoemission data, that were previously used to model
the spin susceptibility \cite{norman2007}.  The coefficients of these dispersions
(tb1, tb2, and tb4) can be found in that work.  tb1 is based on earlier photoemission
data for Bi2212 \cite{tb1} and is characterized  by a van Hove singularity at $(\pi,0)$ that
is 34 meV below the Fermi energy, resulting in a sizable anisotropy of the Fermi velocity around
the Fermi surface.  tb2 is based on more recent photoemission data for Bi2212 \cite{tb2} which indicates
an isotropic Fermi velocity.  tb4 is based on photoemission data for underdoped 
La$_{2-x}$Sr$_x$CuO$_4$ (LSCO) \cite{tb4}.

In Fig.~\ref{Fig:colormap_chi}, we show $\chi(\bq,\Omega=0)$.
There are two sets of maxima.  One set forms a box-like structure around the zone center,
and it is this set which will be assumed to dominate the charge response since we anticipate
that $g_{\bq}$ will be maximal at $q=0$.  The second set is concentrated 
around the $(\pi,\pi)$ point, and is thought to dominate the
spin response since the superexchange interaction is maximal at this point.
For the first set, the absolute maximum of $\chi$ typically occurs along the zone diagonal
($q_x=q_y$).  This can be qualitatively understood by employing an argument due to
Schulz \cite{schulz}.  Although a bond aligned vector does a better job  of nesting
the antinodal sections of the Fermi surface near $(\pi,0)$, a diagonal vector connects
twice as many Fermi surface faces.
Therefore, were it not for the momentum dependence of the interaction, one
would conclude that the CDW typically cannot be directed along the bond directions,
at least within the model considered here.

\begin{figure}
\includegraphics[width=\columnwidth,clip]{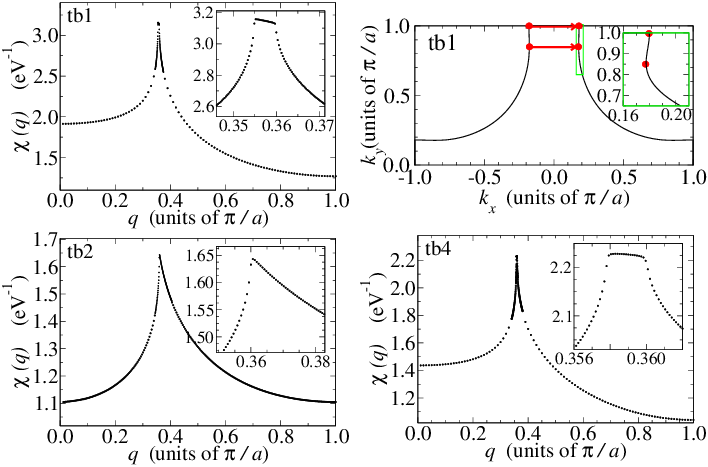}
\caption{\label{Fig:chi_T_0} (Color online) Susceptibility for $T=0$ along $\Gamma-X$ for 
dispersions tb1
(top left), tb2 (bottom left) and tb4 (bottom right). In all cases, $\chi$
is finite due to imperfect nesting. 
The Fermi surface for tb1 is shown in the top right panel.  For dispersions
tb1 and tb4, the susceptibility has two cusps, corresponding to momenta connecting
points in a quadrant of the zone (shown for tb1 by the dots) with their partners in the adjacent quadrant.
The characteristic bending of the Fermi surface near the antinodal direction,
which is similar for the tb1 and tb4 fits, is absent for the tb2
dispersion where the susceptibility has a single cusp. In all cases, on one
side of each cusp at $Q_{x0}$, the slope of  $\chi(Q_x,0)$ is finite, while on the other side,
$\chi(Q_{x0},0)-\chi(Q_x,0)\propto |Q_x-Q_{x0}|^{1/2}$ and the slope is infinite.
}
\end{figure}

In two dimensions, the susceptibility does not diverge, even at $T=0$,
unless the faces of the Fermi surfaces are perfectly nested.
This is never the case for a realistic dispersion, although the Fermi surface does contain
approximately nested sections that are responsible for the sharp peaks in $\chi$. This is
illustrated in Fig.~\ref{Fig:chi_T_0} for the peak lying along $\Gamma-X$.
The discontinuities of the slope at momenta $\bq_1$ and $\bq_2$ (which define
the plateau in the insets) are due to
the points of the Fermi surface $\bk$ and $\bk'$ separated by $\bq_i$ such
that the tangents to the Fermi surface at $\bk$ and $\bk'$ are  
parallel. The lack of divergence in the susceptibility implies that
the charge-density wave can develop only if the interaction strength
exceeds a critical value.

\begin{figure}
\includegraphics[width=0.4\textwidth,clip]{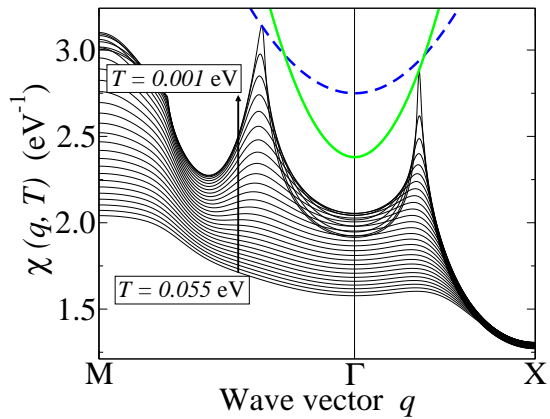}
\caption{\label{Fig:chi_line_tb1_together} (Color online) Depending on the momentum dependence of the
interaction $g_{\bq}$, a CDW with modulations along the Cu-O bonds ($\Gamma-X$) can be
stabilized despite the fact that the maximum of the susceptibility along this direction is
smaller than the peak value along $\Gamma-M$. Thin solid lines describe the
momentum dependence of susceptibility for the tb1 dispersion along the contour $M-\Gamma-X$,
for temperatures from $T=0.001$ eV to  $T=0.055$ eV, with increments
of  $T=0.002$ eV. The inverse interaction strength  $-1/2g_{\bq} = \alpha + \beta (a\bq)^2$
for two different sets of parameters $(\alpha,\beta)$ is shown by
solid green  ($\alpha = 2.38$ eV$^{-1}$, $\beta= 0.40$ eV$^{-1}$)
and dashed blue ($\alpha = 2.75$ eV$^{-1}$, $\beta = 0.15$ eV$^{-1}$) parabolas.}
\end{figure}

The  ordering vector in cuprates observed by scanning tunneling microscopy
lies along the Cu-O bonds, in contrast with our results for $\chi$.  On the other hand,
the true ordering vector is determined by the product of $\chi$ and $g$.
We anticipate that $g_{\bq}$ has a maximum at $q=0$ and falls off smoothly with increasing $q$.  
If this fall off is relatively isotropic and steep enough,
then, as demonstrated in Fig.~\ref{Fig:chi_line_tb1_together}, this can lead to ordering
along the bond direction instead, since $q$ along the diagonal of the box structure 
shown in Fig.~\ref{Fig:colormap_chi} is $\sqrt{2}$ times larger than $q$ along  the bond direction.
In general, the momentum $\bq=\bQ$ at which the charge-density instability
occurs is determined by the highest temperature such that the condition
\begin{equation}
\label{instability_equation}
-\frac1{2g_{\bq}} = \chi(\bq, T_c)
\end{equation}
is satisfied.  
For concreteness, we assume that the interaction has the following
simplified form:
\begin{equation}
-\frac1{2g_{\bq}} \approx \alpha+\beta |a\bq|^2
\end{equation}
As this small $q$ expansion is most sensitive to the longer range part of $g$, our simple
approximation cannot be directly mapped onto the extended Hubbard model mentioned
in Section II.

\begin{figure}
\includegraphics[width=0.5\textwidth,clip]{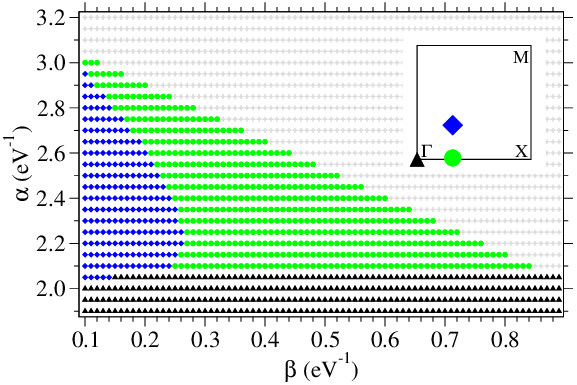}\hfil%
\caption{\label{Fig:instability_diagram} (Color online) Instability diagram for the tb1
dispersion. The  region marked by gray ``$+$'' symbols corresponds to interactions so weak that even
at $T=0$, no CDW ordering occurs. For $(\alpha, \beta)$ in the
parts of the diagram  marked by green circles (blue diamonds), the instability happens at a momentum
lying along $\Gamma-X$ ($\Gamma-M$).  Even for moderately localized  interactions
($\beta$ small)
the instability first appears at $\bq=0$, provided that the interaction
is sufficiently strong  ($\alpha \lesssim 2$ eV$^{-1}$).
}
\end{figure}

Given this form of the interaction, the first task is to identify the
momentum $\bQ$ and the temperature $T_c$ at which the CDW first develops.
As shown in Fig.~\ref{Fig:instability_diagram},
the solutions generally fall in four different classes. For a fixed
curvature $\beta$, there are no solutions provided that $\alpha$ is 
large enough. This was to be expected, as the logarithmic divergence
of the susceptibility is cut off due to imperfect nesting. Thus, 
for sufficiently weak interactions, no ordering occurs even at $T=0$.
In the opposite limit of strong interactions and medium- to 
long-range  interactions, the instability is at $q=0$.
This is due to the strong reduction in the momentum dependence of $\chi$
as the temperature is raised, as can be appreciated from Fig.~\ref{Fig:chi_line_tb1_together}.
In this high $T$ limit, the susceptibility maxima are confined to the region near the $M$ point
of the zone, which leads to solutions for these wave vectors for small values of the curvature
$\beta \lesssim 0.1$.  This is not shown in
Fig.~\ref{Fig:instability_diagram}.  The reason is that the parabolic
approximation we apply for $1/g_{\bq}$ is only valid near $q=0$.
We anticipate that for larger values of $q$, the charge interaction is suppressed,
and the magnetic interactions become dominant instead.

We therefore focus on the more pertinent cases of 
the two regions that  are shown in Fig.~\ref{Fig:instability_diagram} by
the green circles and blue diamonds. In these cases,
the instability occurs either at a momentum along $\Gamma-X$ (green circles)
or $\Gamma-M$ (blue diamonds); two representative examples are shown in Fig.~\ref{Fig:chi_line_tb1_together}
(solid green and dashed blue parabolas).
In either case, the ordering momentum lies in the immediate vicinity of
the ``box'' structure that surrounds the $\Gamma$-point
(Fig.~\ref{Fig:colormap_chi}).  It is straightforward to show that 
the boundary separating the regions shown by the green circles and the
gray ``+'' signs in the parameter space must be a straight line. This boundary
corresponds to the limit $T_c\to 0$; 
in Fig.~\ref{Fig:chi_line_tb1_together}  it
represents a family  of parabolas with different $\alpha$ and $\beta$, all passing
through the maximum  of the zero-temperature susceptibility at a
momentum $Q_0\approx 0.355\pi/a$ (Fig.~\ref{Fig:chi_T_0}, tb1). The boundary is thus described by a straight line:
\begin{equation}
\alpha + \beta a^2 Q_0^2 = \chi(T=0, Q_0)
\end{equation}
The other  main features of the instability diagram can be understood in a
similar fashion.

\section{Quartic terms and the symmetry of the CDW in cuprates}

\begin{figure}
\includegraphics[width=0.5\textwidth,clip]{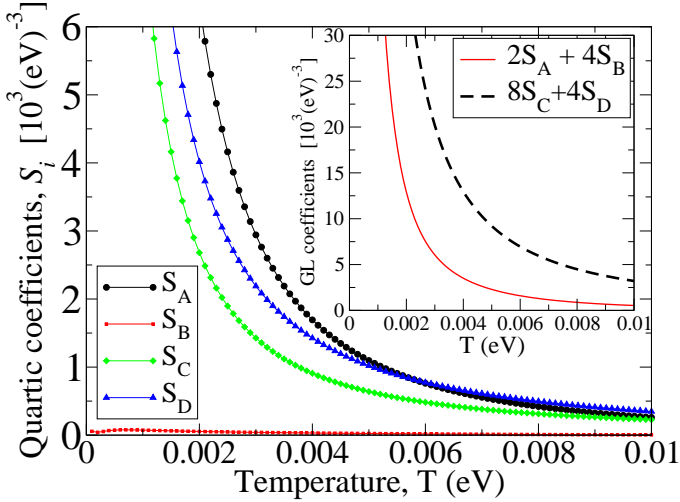}\hfil%
\caption{\label{Fig:S4coeff_T_dependence} (Color online) Temperature dependence of the
quartic coefficients $S_A$,  $S_B$,  $S_C$,  and $S_D$ for $\bQ
\approx (0.357 \pi/a$, 0) (tb1 dispersion). The inset shows the linear combinations of 
the quartic coefficients that determine the symmetry of the charge
density wave. The condition $\gamma = (8S_C+4S_D)/(2S_A+4S_B) > 1$, which
is fulfilled for all temperatures in this case, corresponds to one-dimensional stripes.
}
\end{figure}

Once the ordering momentum $\bQ$ is known,  the coefficients
of the Ginzburg-Landau free energy, evaluated at the transition
temperature $T=T_c$, can be computed rather easily.  
Fig.~\ref{Fig:S4coeff_T_dependence} illustrates the temperature
dependence of the quartic coefficients calculated for the momentum
$\bQ = Q_0\hx$, where $Q_0 \approx 0.357\pi/a$ is the instability 
momentum for a representative case of $\alpha = 2.4$ eV$^{-1}$ and
$\beta = 0.38$ eV$^{-1}$.  At high temperatures ($T \gg t$), the
leading-order result for all coefficients $S_i$ ($i=A, B, C, D$) has the same functional
form  $48 T^{-3} + {\ldots}$, as can be shown by a straightforward
expansion of the integrand in Eq.~(\ref{S_4_summed_generalcase}) in powers
of $\xi_k/T$. In the opposite limit of low
temperatures, the coefficients $S_A$, $S_C$, and $S_D$ diverge as $T^{-3/2}$, while
$S_B$ appears to remain finite.  At high temperatures, $\gamma \to 2$,
independent of any particular form of the dispersion.
As can be seen from Fig.~\ref{Fig:S4coeff_T_dependence},
$\gamma>1$ is satisfied at all temperatures. 
Therefore, in the simplified description where the additional harmonics
are neglected, one would conclude that 1D stripe ordering 
at a momentum $Q_0$ is the preferred state. The effect of the harmonics
will be discussed next.

Notice that while the coefficients of the quartic term in Eq.~(\ref{Seff})
are nominally determined only by the dispersion $\xi_{\bk}$ (Eqs.~(\ref{S_4_summed_generalcase})
and (\ref{SABCD})), they also indirectly depend on the interaction, $g_{\bq}$, since this enters into the determination of
the ordering vector as discussed above.
However, the functional form of the interaction affects quartic coefficients
in another fashion, which turns out to be rather pronounced.  When
deriving $F(\Delta)$,  we left out of the final expression Eq.~(\ref{Seff})
all momenta but those where the instability first develops. It happens that the
inclusion of the third-order terms in the effective action involves harmonics of the form
$\Delta_{2\bQ}$  and $\Delta_{\pm\bQ\pm\barQ}$.  These terms lead to a 
renormalization of the fourth order coefficient and, for a wide range of parameters,
this can affect the preferred symmetry of the CDW, as shown in
Ref.~\onlinecite{hongkivelson} for the case of the rare earth tri-tellurides.
That is, inclusion of these harmonics leads to an additional
contribution to $F(\Delta_{\bQ}, \Delta_{\barQ})$ of the form

\begin{multline}
\delta F_3(\Delta_{\bQ, \barQ}) =\\
r_{2\bQ} (|\Delta_{2\bQ}|^2+|\Delta_{2\barQ}|^2)
+r_{\bQ+\barQ} (|\Delta_{\bQ+\barQ}|^2+|\Delta_{\bQ-\barQ}|^2)
\\
+b_{\bQ}\left(\Delta_{\bQ}^2\Delta_{-2\bQ}  + \Delta_{\barQ}^2\Delta_{-2\barQ} 
+ c.c.\right)\\
+c_{\bQ}\left(\Delta_{-\bQ-\barQ}\Delta_{\bQ}\Delta_{\barQ}+ 
\Delta_{\barQ-\bQ} \Delta_{\bQ}\Delta_{ -\barQ} +c.c.\right)
\end{multline}
where
\begin{align}
b_{\bQ} &= -2 S_3(\bQ, \bQ, -2\bQ)\\
c_{\bQ} &= -4 S_3(\bQ, \barQ, -\bQ-\barQ)
\end{align}

\begin{figure}
\includegraphics[width=\columnwidth,clip]{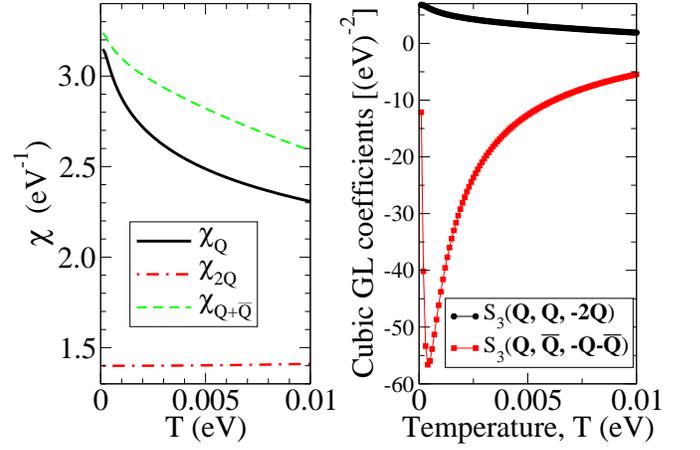}\hfil%
\caption{\label{Fig:S3coeffs} (Color online) Left panel: Temperature
dependence of the susceptibility at momenta $\bQ$, $2\bQ$, and $\bQ+\barQ$
for momentum $\bQ \approx (0.357 \pi/a, 0)$ (tb1 dispersion).
Right panel: Temperature dependence of the
third-order coefficients $S_3(\bQ,\bQ, -2\bQ)$ and $S_3(\bQ, \barQ,
-\bQ- \barQ)$.
}
\end{figure}

In the expression for $\delta F_3$ above, we omitted all terms higher than cubic order.
These harmonics can be integrated out, as was shown by Yao
\textit{et al.} \cite{hongkivelson}, and the resulting
correction reads
\begin{equation}
\delta F_3  = -\frac{b_{\bQ}^2}{r_{2\bQ}} (|\Delta_{\bQ}|^4
+ |\Delta_{\barQ}|^4) - 2\frac{c^2_{\bQ}}{r_{\bQ+\barQ}}
|\Delta_{\bQ}|^2|\Delta_{\barQ}|^2
\end{equation}
These terms depend on the specific form of the interaction $g_{\bq}$,
which affects the coefficients $r_{\bQ}$ and the choice of momentum $\bQ$.
The coefficients $S_3$ are shown in Fig.~\ref{Fig:S3coeffs}.

\begin{figure*}
\includegraphics[width=\textwidth,clip]{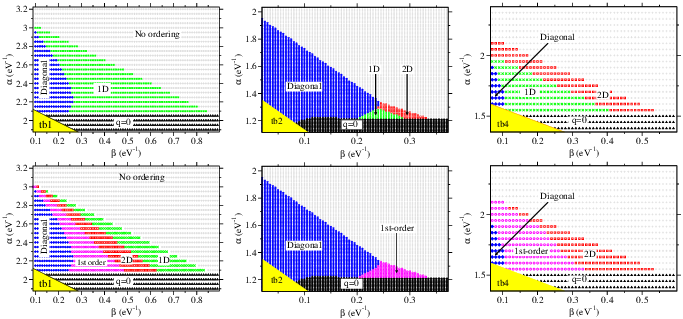}\hfil%
\caption{\label{Fig:refinedinstability_diagram} (Color online) A refined version 
of the instability diagram of Fig.~\ref{Fig:instability_diagram}. The
top (bottom) row corresponds to the instability diagrams
without (with) cubic corrections. 
Instability at momenta lying along the Cu-O bonds can result in a 1D stripe phase
(green crosses) or a 2D ``checkerboard'' phase (red squares).
In the region marked by magenta circles, the cubic corrections that
involve modes with momenta $\bQ + \barQ$ destabilize the GL free energy
(at the quartic level) by violating the stability condition  $\gamma>-1$,
corresponding to a first-order transition to a state with finite 
$\Delta_{\bQ}$, $\Delta_{\barQ}$, $\Delta_{\bQ+\barQ}$, and higher
harmonics.  The yellow shaded region (lower left corner) of each diagram
indicates a regime where the transition temperature - formally obtained
as a solution of Eq.~(12) - is so high, the susceptibility peaks
are smeared, and our simple GL approach should not be applied.
}
\end{figure*}

One can estimate the effect of the
corrections by noticing that the GL coefficients are modified as follows:
\begin{align}
2S_A+4 S_B  &\to 2S_A+ 4 S_B +\frac{ 4
S_3(\bQ, \bQ, -2\bQ)^2}{\displaystyle
\frac1{2g_{2\bQ}}+\chi_{2\bQ}}\label{renormAB}\\
8S_C+4 S_D &\to 8S_C+4 S_D +\frac{16
S_3(\bQ, \barQ, -\bQ-\barQ) ^2}{\displaystyle \frac1{2g_{\bQ+\barQ}}+\chi_{\bQ+\barQ}}\label{renormCD}
\end{align}
In the situation shown in Fig.~\ref{Fig:chi_line_tb1_together},
the additional term in Eq.~(\ref{renormAB})
can be neglected (i.e., the response at $2\bQ$ is typically small).
On the other hand, the one in Eq.~(\ref{renormCD})
can be significant, given the box-like structure of $\chi$
about the zone center  (i.e., the response at $\bQ + \barQ$ can be large).
As a result, the net effect of the
cubic corrections on the GL free energy is a reduction of the coefficient
$\gamma$ describing the relative magnitude of the mixed term 
$|\Delta_{\bQ}|^2|\Delta_{\barQ}|^2$. Consequently, the range of parameters for
which the checkerboard symmetry is realized could be generally
increased.  Evaluation of the renormalized quartic coefficients from
Eqs.~(\ref{renormAB}) and (\ref{renormCD}) leads to a refined version of 
the instability diagram shown in Fig.~\ref{Fig:refinedinstability_diagram}.

\textit {Dispersion tb1}. In the region of the diagram that corresponds to low transition
temperatures, the contribution from the quartic coefficients dominates
over the third-order terms, and the preferable state is one-dimensional.
For smaller curvature $\beta$, as one approaches the part of the
diagram where the instability occurs along a diagonal wave vector,
the contribution of the third-order terms becomes
more pronounced and results in a transition to a two-dimensional
state, as described above. Both the one-dimensional and the
two-dimensional cases are characterized by the dominant order parameter
$\Delta_{\bQ}\propto (T_c-T)^{1/2}$. In all cases, there are
subdominant higher harmonics of the order parameter at wave vectors
$2\bQ$, $\bQ + \barQ$, etc.~that in the vicinity of $T_c$ behave
as integer powers of $\Delta_{\bQ}$.  For example, the most pronounced 
subdominant order parameter for the ``checkerboard'' state is $\Delta_{\bQ+\barQ} \propto
|\Delta_{\bQ}|^2/r_{\bQ+\barQ} \propto (T_c-T)/r_{\bQ+\barQ}$.
Note that as $\beta$ is reduced, $r_{\bQ +\barQ}$ also decreases
(Fig.~\ref{Fig:chi_line_tb1_together}). Thus for fixed $T_c-T$, 
the subdominant order $\Delta_{\bQ+\barQ}$ will be growing, and one
expects that the simple GL scenario considered so far will break down
at some point.

Indeed, for even lower values of $\beta$, 
the renormalized value of  $\gamma$ falls into the  $\gamma < -1$ range.
In other words, the third-order terms destabilize our reduced GL free energy expression that 
included only  terms up to the fourth order in $\Delta_{\bQ}$.
This instability implies the necessity to include 
higher-order terms in the GL free energy expression for
that part of the parameter space. These terms would generally 
restore the stability of the GL free energy, but would result
in a  first-order transition that cannot be described in any quantitative way using our approach.
At  the transition temperature $T_c$, a finite amplitude of $\Delta_{\bQ}$
develops, and unless a weakly first order transition occurs, one in 
principle has to include an infinite number of terms in the GL free energy 
to describe it. The coupling of the order parameter $\Delta_{\bQ}$ to the modes
at $2\bQ$, $\bQ + \barQ$, and higher harmonics in this case implies that
$\Delta_{2\bQ}$, $\Delta_{\bQ+ \barQ}$, etc.~also acquire finite values
just below $T_c$. Since none of the higher harmonics have been 
observed in experiment, a detailed analysis of this phase is 
left for future study.

\textit {Dispersion tb2}. The phase diagram for tb2 has significant
differences from the tb1 case.  The region of stability for diagonal
ordering is enhanced at the expense of the 1D state, and in addition
the 2D checkerboard state now appears.
But the cubic corrections in this case are rather large, and with their
inclusion, over the entire range of parameters 
where 1D and 2D second-order transitions would be expected,
first-order transitions occur instead.

\textit {Dispersion tb4}. The tb4 case looks more similar to the tb1
case than the tb2 one, except that 2D order is now present over
a region of parameter space where order first appears.  Inclusion 
of the cubic corrections leads to a complete suppression of the 1D state
in favor of a first-order transition,
but the checkerboard order remains stable.

\section{Conclusions}
One faces two difficulties when attempting to reconcile the
atomic-scale modulations  observed in real space
by scanning tunneling microscopy (STM) with the energy
dispersions obtained from photoemission.  First, one must account for the fact
that the modulation wave vector observed by STM is directed
along the Cu-O bonds, while the charge susceptibility extracted
from the photoemission dispersions is largest along the diagonal of the
zone.  At the level of a weak-coupling
theory with an effective electron-electron interaction,
this apparent contradiction implies that the interaction must be 
momentum dependent. Moreover, one obtains rather stringent 
constraints on the range and strength of this effective interaction.

While both LSCO and Bi2212 have similar dispersions, the differing 
ordering tendencies 
observed in these materials should not come as a surprise, since
small differences in the dispersions and the interaction coefficients
$\alpha$ and $\beta$ are sufficient to move one from the 1D to 2D regions of
the phase diagram.   In all cases, one obtains a qualitatively similar instability
diagram at the level of the susceptibility analysis, i.e., diagrams that 
determine whether the instability momentum lies along
the Cu-O bonds or diagonally. 

Even when the interaction is such that the instability momentum 
is oriented correctly, there remains the question of a single $Q$ (stripe)
versus a double $Q$ (checkerboard) state.
This is where the results diverge sharply, even for materials 
with seemingly similar charge susceptibilities.
As the quartic terms are sharply peaked functions of momentum
near the ``box'' where the susceptibility is
peaked, the detailed phase diagram that determines the
energetically preferred state is a sensitive function of the dispersion.

Interestingly, the well known difficulty of
differentiating whether the observed modulation patterns
are uni-directional or checkerboard-like implies that although uni-directional
behavior is the most likely, the real materials are 
near the 1D-2D phase boundary \cite{robertson} in Fig.~8.
It is also interesting to note that the cubic corrections tend to destabilize
1D order in favor of 2D order.  In that context, recent observations in underdoped
YBa$_2$Cu$_3$O$_{6+x}$ have seen 2D order \cite{ghiringhelli,chang} and thus
our work should be of interest in this regard.

\acknowledgments
We would like to thank  S. Davis, O. Vafek and H. Yao for discussions
and correspondence. This work was supported by the Materials Sciences and Engineering
Division, Basic Energy Sciences, Office of Science, US DOE.

\end{document}